\documentstyle[12pt,epsf]{article}
\makeatletter
\@addtoreset{equation}{section}
\makeatother

\def\xib{\bar{\xi}}

\begin{document}

\begin{titlepage}

\def\thefootnote{\fnsymbol{footnote}}
\begin{flushleft}
ITP-UH-06/94 \hfill April~1994\\
ITP-SB-94-16
cond-mat/9404091
\end{flushleft}

\vspace*{\fill}

\begin{center}
{\Large \bf Differential equation for a correlation function\\
        of the spin-${1\over2}$ Heisenberg chain}
\vfill

{\sc Holger Frahm}\footnote{%
e-mail: {\tt frahm@itp.uni-hannover.de}}\\
{\sl Institut f\"ur Theoretische Physik\\[-6pt]
     Universit\"at Hannover\\[-6pt]
     D--30167~Hannover, Germany}\\[8pt]
{\sc Alexander R.\ Its}\footnote{%
e-mail: {\tt itsa@math.iupui.edu}}\\
{\sl Dept. of Mathematical Sciences\\[-6pt]
     Indiana University-Purdue University at Indianapolis (IUPUI)\\[-6pt]
     Indianapolis, IN 46202--3216, U. S. A.}\\[8pt]
{\sc Vladimir E.\ Korepin}\footnote{%
e-mail: {\tt korepin@max.physics.sunysb.edu}}\\
{\sl Institute for Theoretical Physics\\[-6pt]
     State University of New York at Stony Brook\\ [-6pt]
     Stony Brook, NY 11794--3840, U. S. A.}\\[8pt]
\vfill
ABSTRACT
\end{center}

\begin{quote}
We consider the probability to find a string of $x$ adjacent parallel spins
in the antiferromagnetic ground state of the model (in a magnetic
field). We derive a system of integro-difference equations which define
this probability. This system is completely integrable, it has Lax
representation and a corresponding Riemann-Hilbert problem. The quantum
correlation funtion is a $\tau$-function of this system.
\end{quote}

\vfill

\end{titlepage}
\setcounter{footnote}{0}

\section{Introduction}
In this paper we continue to investigate a special correlation function in
the XXX Heisenberg model. The model is defined in terms of the hamiltonian
\begin{equation}
   {\cal H} = \sum_{j} \sigma_j^x \sigma_{j+1}^x
                     + \sigma_j^y \sigma_{j+1}^y
                     + \sigma_j^z \sigma_{j+1}^z
                     - h \sigma_j^z\ .
   \label{xxx}
\end{equation}
where the sum is over all integers $j$, $\sigma^\alpha$ are Pauli matrices
and $h$ is an external magnetic field. This model was solved by Bethe
\cite{bethe:31} by explicit construction of eigenfunctions of the
model. The corresponding Bethe Ansatz equations define the allowed spectrum
of excitations over the ferromagnetic vacuum $\vert \uparrow \uparrow
\uparrow \ldots \rangle$ which are characterized by complex valued
rapidities $\lambda$.  The {\em antiferromagnetic} ground state $\vert 0
\rangle$ is obtained by filling the sea of these excitations with real
$\lambda$. The case of finite magnetic field $h$ leading to a nonzero
magnetization of the ground state was first studied by Griffiths
\cite{grif:64}. The energy of the excitiations above this ground state is
given in terms of the solution of the following integral equation:
\begin{equation}
   \varepsilon(\lambda)
   + {1\over2\pi} \int^\Lambda_{-\Lambda} d\mu
            K(\lambda, \mu) \varepsilon(\mu ) d \mu =
   2h - {2\over \lambda^2+{1\over 4}}\ , \qquad
   \varepsilon(\pm\Lambda)=0\ .
   \label{dresse}
\end{equation}
The kernel of this integral equation is $K(\lambda,\mu) = {2/\left(1+
(\lambda - \mu)^2\right)}$. The boundaries $\Lambda$ of integration are a
function of the magnetic field $h$ through the requirement of vanishing
$\varepsilon$ at these points. $2\Lambda$ is the size of the Fermi sphere
in this model (in relativistic field theory it can be compared with
cut-off). While the spectrum and thermodynamic properties of the model have
been extracted from the Bethe Ansatz integral equations such as
(\ref{dresse}) correlation functions have not been computed: The reason for
this is the complicated form of Bethe's original wave functions. However,
within the Quantum Inverse Scattering method the Bethe Ansatz states are
constructed within an algebraic framework which has been used successfully
to compute correlation functions in systems such as the nonlinear
Schr\"odinger model \cite{vladb}. Within this approach three main steps
have to be performed:

\begin{itemize}
\item
First, starting from the solution of the model by means of the algebraic
Bethe Ansatz, the correlation functions have to be expressed as
determinants of Fredholm integral operators, see formula (\ref{fred}).

\item
Then a description of these determinants in terms of the solutions of
certain integrable integro-difference equations has to be found. These
equations determine completely the correlation functions in question, see
(\ref{result:f}), (\ref{B:dpsi}). Fredholm determinant can be identified as
a fredholm determinant of the Gelfand-Levitan-Marchenko integral operator
of these equations evaluated on special solution (for definition of
Gelfand-Levitan-Marchenko integral operator see
\cite{fata:book,ablowitz+segur}).

\item
The long distance asymptotics of the correlatiors can be obtained from the
solution of a matrix Riemann-Hilbert problem related to these
integro-difference equations. The Riemann-Hilbert problem also provides an
adequate parametrization of the space of the solutions for
integro-difference equations. It sould be noted that just this type of
parametrization is most suitable for the asymptotic analysis of the
integrable systems (see review paper \cite{deiftx:93}).
\end{itemize}
This means that quantum correlation function is a $\tau$ function (in a
sense of \cite{jimbo:80}) of these integro-difference equations. In order
to define the correlation function that we are interested in we introduce
the operators
\begin{equation}
   P_j = {1\over2} (\sigma_j^z + 1)
       = \left( \begin{array}{cc} 1 & 0 \\ 0 & 0
   \end{array} \right)\ ,
\end{equation}
projecting onto the state with spin up at site number $j$. In terms of
these operators the correlation function that we study here is
\begin{equation}
P(x) = \langle 0 |\; \prod_{j=1}^x P_j \;| 0 \rangle .
\end{equation}
The physical meaning of $P(x)$ is the probability of finding a string of
$x$ (adjacent) parallel spins up (i.e.\ a piece of the {\em ferro}magnetic
ground state) in the {\em antiferro}magnetic ground state $|0\rangle$ of
the model (\ref{xxx}) for a given value of the magnetic field $h$. {}From a
mathematical point of view this correlator turns out the simplest one to be
considered (see \cite{vladb}).

Recently, the first step in the program outlined above has been solved,
i.e.\ an expression of the function $P(x)$ in terms of a Fredholm
determinant of an integral operator \cite{korx:94} has been constructed:
\begin{equation}
   P(x) = { { \left( 0 \left|
                  \det \left(1 + \widehat{V}^{(x)} \right)
                       \right| 0 \right) }\over
            { \det \left( 1 + \widehat{K}/{2\pi} \right) }}\ .
   \label{fred}
\end{equation}
Here $\widehat{K}$ is the integral operator appearing in (\ref{dresse}),
$\widehat{V}^{(x)}$ is another integral operator acting on a function
$f(\lambda)$ as
\begin{equation}
   \left( \widehat{V}^{(x)} f \right) (\lambda)
          = \int_{-\Lambda}^{\Lambda} d\mu\ V^{(x)}(\lambda|\mu) f(\mu)\ .
\end{equation}
Its kernel $V$ is given by
\begin{equation}
   V^{(x)}(\lambda|\mu) = {1\over2\pi} \left(
       {e^{(x)}_+(\lambda) e^{(x)}_-(\mu)
           \over (\lambda - \mu)(\lambda - \mu +i)}
     + {e^{(x)}_-(\lambda) e^{(x)}_+(\mu)
           \over (\mu - \lambda)(\mu - \lambda +i)}
     \right)
   \label{vkernell}
\end{equation}
with functions $e_\pm^{(x)}$
\begin{equation}
   e^{(x)}_\pm(\lambda) = \left(
       \left({2\lambda+i\over2\lambda-i}\right)^x e^{\phi(\lambda)}
                          \right)^{\pm{1\over2}}\ .
\end{equation}
The appearence of the dual quantum fields $\phi(\lambda)$ is a consequence
of the interacting nature of the Heisenberg model (see Section XI.1 of
\cite{vladb}). They are acting in the bosonic Fock space with vacuum $|0)$.
This means that $\det(1+\widehat{V})$ will be a complicated operator in
this Fock space. However, by construction the operators $\phi(\lambda)$
commute among each other:
\[
   \left[\phi(\lambda)  ,\phi(\mu) \right] = 0\ .
\]
In fact they are related to canonical Bose fields $a$ and
$a^\dagger$ which standard commutation relations
\begin{eqnarray}
  && \bigg[ a(\lambda), a^\dagger (\mu) \bigg] = \delta(\lambda - \mu)
     \nonumber\\
  && \bigg[ a(\lambda) , a (\mu) \bigg] =
     \bigg[ a^\dagger (\lambda),a^\dagger (\mu) \bigg] = 0\ .
\end{eqnarray}
These fields annihilate the Fock vacuum and its dual, resp.:
$a(\lambda)|0)=0=(0|a^\dagger(\lambda)$. The dual fields $\phi$ are given
by the expressions:
\begin{equation}
   \phi (\lambda) = a(\lambda)
     - \int^\infty_{-\infty} d\nu \ln \bigg[ 1 +(\lambda-\nu)^2 \bigg]
            a^\dagger(\nu)\ .
\end{equation}

The plan of this paper is as follows: in the following section we shall
present the main results, namely the complete system of
integro-difference equations that drive the Fredholm determinant of the
operator $\widehat{V}$ in the correlation function $P(x)$ (\ref{fred}). In
Section~3 we will give the derivation of these results and in Section~4 the
Riemann-Hilbert problem associated with the equations is studied.

We also want to mention that quantum correlation function were first
described by differential equation [Painlev\'e] by Barouch, McCoy, Wu and
Tracy and McCoy \cite{barouch:73,trmc:73}.

\section{Main results\label{sec:results}}
We obtain a complete set of integro-difference equations for eight
functions $F_\pm^{(x)}(s)$, $G_\pm^{(x)}(s)$ and the kernels of integral
operators $B_{\pm,\pm}^{(x)}(s,t)$ as well as the Riemann-Hilbert problem
machinery (see e.g.\ \cite{fata:book}) to integrate these equations. In
terms of these quantities we can compute the Fredholm determinant of
$1+\widehat{V}$ using the relations
\begin{equation}
   { {\det\left( 1+ \widehat{V}^{(x+1)} \right)} \over
     {\det\left( 1+ \widehat{V}^{(x)} \right)} }
   = \det\left( 1 + B_{-+}^{(x)} \right)
   \label{vvd1}
\end{equation}
(which is the lattice equivalent of the logarithmic derivative of
$\det\left( 1+\widehat{V}^{(x)} \right)$) and
\begin{equation}
     { {\det\left( 1+ \widehat{V}^{(x+1)} \right)} \over
       {\det\left( 1+ \widehat{V}^{(x)} \right)} } \;
     { {\det\left( 1+ \widehat{V}^{(x-1)} \right)} \over
       {\det\left( 1+ \widehat{V}^{(x)} \right)} } =
     \det \left( 1 - \left(B_{++}^{(x)}\right)^\top B_{--}^{(x)} \right)\, .
   \label{vvd2}
\end{equation}
($(B)^\top$ is the transpose of $B$). In addition we have
\begin{eqnarray}
   -i \partial_\psi \ln \det\left( 1+\widehat{V} \right) &=& {1\over 2\pi}
     \int_s \Bigl\{F_+(s) \partial_\psi F_-(s)
                 - F_-(s) \partial_\psi F_+(s)
       \nonumber \\
  &&\qquad \qquad
                 + G_-(s) \partial_\psi G_+(s)
                 - G_+(s) \partial_\psi G_-(s)
            \Bigr\} \nonumber \\
  && + {1\over 4\pi^2} \; {\xi+\xib \over \xi-\xib} \;
        \left( \int_s F_+(s) G_-(s)
                    - F_-(s) G_+(s) \right)^2 \; .
  \label{eq:vvdpsi}
\end{eqnarray}
Here the integration with respect to $s$ is to be performed with the
measure
\begin{equation}
   \int_s f(s) = \int_0^\infty ds e^{-s} f(s)\ .
\end{equation}
The variable $\psi$ is related to the cut-off $\Lambda$ which is a function
of the magnetic field:
\begin{equation}
   e^{i\psi} \equiv \xi = {{2\Lambda-i}\over{2\Lambda+i}} \; .
   \label{def:psi}
\end{equation}
(Note that $\psi<0$).

The integro-difference equations for $f_\pm$ and $B_{\pm,\pm}$ are
\begin{eqnarray}
   {1 \over \sqrt{\xi}} F_+^{(x+1)} &=& {1\over \xi}
        \left\{ F_+^{(x)}
          - \left( 1- B_{+-}^{(x+1)} \right)
            \left( B_{++}^{(x)} \right)^\top F_-^{(x)} \right\}
   \nonumber \\
   {1 \over \sqrt{\xi}} F_-^{(x+1)} &=&
      {1\over \xi} \left\{
        \left( \xi + B_{--}^{(x+1)} \left(B_{++}^{(x)}\right)^\top
        \right) F_-^{(x)}
        - B_{--}^{(x+1)} \left( 1+ \left(B_{+-}^{(x)}\right)^\top
                         \right) F_+^{(x)} \right\}
   \nonumber \\
   \label{result:f} \\
   {1 \over \sqrt{\xib}} G_+^{(x+1)} &=& {1\over \xib}
        \left\{ G_+^{(x)}
          - \left( 1- B_{+-}^{(x+1)} \right)
            \left( B_{++}^{(x)} \right)^\top G_-^{(x)} \right\}
   \nonumber \\
   {1 \over \sqrt{\xib}} G_-^{(x+1)} &=&
      {1\over \xib} \left\{
        \left( \xib + B_{--}^{(x+1)} \left(B_{++}^{(x)}\right)^\top
        \right) G_-^{(x)}
        - B_{--}^{(x+1)} \left( 1+ \left(B_{+-}^{(x)}\right)^\top
                         \right) G_+^{(x)} \right\}
   \nonumber
\end{eqnarray}
and
\begin{eqnarray}
   -i {\partial \over \partial \psi} B_{ab} (s,t) &=&
    {i\over 2\pi} \Biggl\{ 
      F_a(s) \left[ F_b(t)
                      + \left( F_+ B_{-b} - F_- B_{+b} \right) (t)
                 \right] \nonumber \\
   && \qquad +
      G_a(s) \left[ G_b(t)
                      + \left( G_+ B_{-b} - G_- B_{+b} \right) (t)
                 \right] \Biggr\}
   \label{B:dpsi}
\end{eqnarray}
where $a,b=\pm$. In these equations the action of the integral operators
$B_{ab}$ is given by
\begin{equation}
   \left( B_{ab} f \right)(s) = \int_s B_{ab}(s,t) f(t), \qquad
   \left( \left( B_{ab} \right)^\top f \right)(s)
       = \int_s B_{ab}(t,s) f(t)\ .
\end{equation}
These equations are completely integrable, they have Lax representation,
see Eqs.~(\ref{shift:fright}), (\ref{f:dpsi2}) and
(\ref{RH:compatibility}).

The additional restrictions necessary to solve the equations uniquely are
provided by the requirements on analyticity and the asymptotic behaviour of
the solutions of the corresponding linear system (Eqs.\
(\ref{shift:fright}), (\ref{f:dpsi2})), i.e.\ by specification of the data
in the corresponding Riemann-Hilbert problem (see Section
\ref{sec:RH-problem}).
\section{Derivation}
The plan of the Section is the following: (i) We shall introduce functions
$f_\pm^{(x)}(z|s)$ as solutions of linear integral equation (\ref{def:f}).
(ii) We shall prove that these funtions satisfy linear difference equation
(\ref{shift:fright}) and linear differential equation (\ref{f:dpsi2}).
Compatibility of these two equations will lead to the system of nonlinear
eqautions (\ref{result:f}), (\ref{B:dpsi}). The unknown functions which
enter these equations are defined by (\ref{def:B}) and by
\[
   F_\pm^{(x)}(s) = f_\pm^{(x)}(\xi|s), \qquad
   G_\pm^{(x)}(s) = f_\pm^{(x)}(\xib|s)
\]
Equations (\ref{shift:fright}) and (\ref{f:dpsi2}) are Lax representation
for nonlinear equations (\ref{result:f}), (\ref{B:dpsi}).

First we have to rewrite the kernel of $\widehat{V}$ (\ref{vkernell}) such
that it is of the ``standard'' form of integrable integral operators
\cite{vladb}. Using the identity
\begin{equation}
   \int_0^\infty ds\ e^{-(1-ix)s} = {1\over 1-ix}
\end{equation}
we obtain
\begin{equation}
   V^{(x)}(\lambda|\mu) = -{i\over2\pi} \int_s
             {e^{(x)}_+(\lambda|s) e^{(x)}_-(\mu|s)
            - e^{(x)}_-(\lambda|s) e^{(x)}_+(\mu|s)
             \over \lambda - \mu}
   \label{vkernell2}
\end{equation}
with $e_\pm(\lambda|s) = e_\pm(\lambda) e^{\pm i\lambda s}$. Furthermore, we
prefer to express $V$ in terms of the transformed variables
\begin{equation}
   z(\lambda) = {2\lambda-i \over 2\lambda+i}\ .
\end{equation}
After performing a similarity transform which leaves the determinant
unchanged the integral operator
 $\widehat{V}^{(x)}$ acts on a function $f(z)$ as
\begin{equation}
   \left( \widehat{V}^{(x)} f \right) (z_1)
                = \int_C dz_2 V^{(x)}(z_1|z_2) f(z_2)
\end{equation}
where the integration is to be performed along the contour
$C: \alpha \to z=\exp{i\alpha}$ where $-\psi < \alpha < 2\pi+\psi$ (see
Fig.~\ref{fig:contour}) and the kernel $V$ is
given by
\begin{equation}
   V^{(x)}(z_1|z_2) = -{i\over{2\pi}}
           \int_s  {{e_+^{(x)}(z_1|s) e_-^{(x)}(z_2|s)
                   - e_-^{(x)}(z_1|s) e_+^{(x)}(z_2|s)} \over
                    {z_1-z_2}}\ .
   \label{vkernelz}
\end{equation}
Now, the functions $e_\pm^{(x)}$ in (\ref{vkernelz}) are given by
\begin{equation}
   e_\pm^{(x)}(z|s) = \left( z^{-x}
                \exp \left( \phi(z) + s\, {z+1\over z-1} \right)
                \right)^{\pm{1\over2}} \; .
   \label{def:e}
\end{equation}
In the following we shall drop the superscript indicating the
$x$-dependence in equations where all quantities are to be taken with the
same $x$.  Note that $V$ is symmetric and nonsingular at $z_1=z_2$.
The resolvent $\widehat{R}$ of $\widehat{V}$ is defined by
\begin{equation}
   (1+\widehat{V})(1-\widehat{R})=1 \quad \Leftrightarrow \quad
   (1+\widehat{V})\widehat{R} = \widehat{V}
\end{equation}
Its kernel $R(z_1|z_2)$  can be written in a form similar to
Eq.~(\ref{vkernelz}), namely \cite{vladb}
\begin{equation}
   R(z_1|z_2) = -{i\over{2\pi}}
            \int_s {{f_+(z_1|s) f_-(z_2|s) - f_-(z_1|s) f_+(z_2|s)} \over
                    {z_1-z_2}} \; .
  \label{rkernelz}
\end{equation}
Here the functions $f_\pm^{(x)}$ are solutions of the linear integral
equations
\begin{equation}
   f_\pm(z_1|s) + \int_C dz_2 V(z_1|z_2) f_\pm(z_2|s)
         = e_\pm(z_1|s) \; .
   \label{def:f}
\end{equation}
In terms of these functions we introduce the integral operators $B_{ab}$,
$a,b=\pm$ acting as $(B f)(s) = \int_t B(s,t) f(t)$ with the kernel
\begin{equation}
  B_{ab}(s,t) = {i\over2\pi} \int_C {dz\over z} f_a(z|s) e_b(z|t)\, ,
      \qquad a,b = \pm \, .
  \label{def:B}
\end{equation}
\subsection{The difference equations in $x$}
Now we are ready to prove Eq.~(\ref{vvd1}): Starting from the
representation (\ref{vkernelz}) of the kernel $V^{(x)}$ in terms of the
functions $e_\pm^{(x)}$ we can write
\begin{equation}
   z_1^{-{1\over2}}\; V^{(x+1)}(z_1|z_2)\; z_2^{1\over2}
      = V^{(x)}(z_1|z_2)
      + {i\over2\pi}\int_s {e_+^{(x)}(z_1|s) e_-^{(x)}(z_2|s) \over z_1}\ .
   \label{shift:v}
\end{equation}
The modification of $V^{(x+1)}$ amounts to a similarity transform only, so
it follows that
\begin{eqnarray}
   \det\left( 1+ \widehat{V}^{(x+1)} \right) &=&
        \det\left( 1 + \widehat{V}^{(x)}
                   + {i\over2\pi} \int_s
           {e_+^{(x)}(z_1|s) e_-^{(x)}(z_2|s) \over z_1} \right)
   \nonumber \\
   &=& \det\left( 1 + \widehat{V}^{(x)} \right)\;
       \det\left( 1 + {i\over2\pi} \int_s
           {e_+^{(x)}(z_1|s) f_-^{(x)}(z_2|s) \over z_1} \right)
   \label{vvd_ef}
\end{eqnarray}
In appendix \ref{app:logder} we show that this indeed equivalent to
(\ref{vvd1}). Together with the identities derived in
appendix~\ref{app:idAB} this also implies Eq.~(\ref{vvd2}).

Next we shall derive the equations (\ref{result:f}). We start from the
defining integral equation (\ref{def:f}) for the functions
$f_\pm^{(x)}(z|s)$. Using (\ref{shift:v}) we obtain
\begin{eqnarray}
   && {1\over\sqrt{z_1}} f_\pm^{(x+1)}(z_1|s)
      + \int_C dz_2 V^{(x)}(z_1|z_2) {1\over\sqrt{z_2}} f_\pm^{(x+1)}(z_2|s)
   \nonumber \\
   && \qquad =
   {1\over\sqrt{z_1}} e_\pm^{(x+1)}(z_1|s)
   - {1\over z_1} \int_t B_{\pm -}^{(x+1)} (s,t) e_+^{(x)}(z_1|t)\ .
   \label{shift:f0}
\end{eqnarray}
To continue from here we need to express the solution of the integral
equation
\begin{equation}
   \left( \left( 1+\widehat{V} \right) g_\pm \right)(z|s) =
     {1\over z} e_\pm^{(x)}(z|s)
   \label{igl:g}
\end{equation}
in terms of $f_\pm^{(x)}(z|s)$. To do so, we rewrite the integral equation
for $f_\pm$ as
\begin{eqnarray}
   && {1\over z_1} f_\pm(z_1|s)
      + \int_C dz_2 V(z_1|z_2) {1\over z_2} f_\pm(z_2|s)
   \nonumber \\
   && \qquad =
      {1\over z_1} e_\pm(z_1|s)
    - \int_C dz2 \left( {1\over z_1}-{1\over z_2} \right)
                      V(z_1|z_2) f_\pm(z_2|s)
   \nonumber \\
   && \qquad =
      {e_\pm(z_1|s) \over z_1}
    - \int_t \left( B_{\pm -}(s,t) {e_+(z_1|t) \over z_1}
                  - B_{\pm +}(s,t) {e_-(z_1|t) \over z_1} \right)\ .
\end{eqnarray}
These equations can be put in matrix form in the $\pm$-components giving
\begin{equation}
   \left(1+\widehat{V} \right)
     \left( \begin{array}{c}
             {f_+(z) \over z} \\ {f_-(z) \over z}
            \end{array}
     \right) =
   \left( \begin{array}{cc}
              1-B_{+-} & B_{++} \\
               -B_{--} & 1+B_{-+}
          \end{array}
    \right)
       \left( \begin{array}{c}
             {e_+(z) \over z} \\ {e_-(z) \over z}
            \end{array}
     \right)\ .
\end{equation}
To proceed we introduce operators $A_{ab}$ that satisfy
\begin{equation}
   \left( \begin{array}{cc}
              1-B_{+-} & B_{++} \\
               -B_{--} & 1+B_{-+}
          \end{array}
    \right)^{-1} =
    \left( \begin{array}{cc}
              A_{+-} & A_{++} \\
              A_{--} & A_{-+}
          \end{array}
    \right)\ .
   \label{def:A}
\end{equation}
In terms of these new operators we obtain the solution of the integral
equation (\ref{igl:g}) for $g_{\pm}$ in terms of $f_\pm$:
\begin{eqnarray}
   g_+(z|s) &=& {1\over z} \left(A_{+-} f_+ + A_{++} f_-\right)(z|s)
    \nonumber \\
   g_-(z|s) &=& {1\over z} \left(A_{--} f_+ + A_{-+} f_-\right)(z|s)\ .
   \label{sol:g}
\end{eqnarray}
Note that as $B_{ab}$ the operators $A_{ab}$ act on the $s$ dependence only
and leave $z$ untouched. This result allows to rewrite (\ref{shift:f0})
giving a set of integro-difference equations for the functions $f_\pm$
\begin{eqnarray}
   {1\over\sqrt{z}} f_+^{(x+1)}(z|s)  &=& {1\over z}
     \left( 1-B_{+-}^{(x+1)}\right)
      \left( A_{+-}^{(x)} f_+^{(x)} + A_{++}^{(x)} f_-^{(x)} \right) (z|s)\ ,
   \nonumber \\
   {1\over\sqrt{z}} f_-^{(x+1)}(z|s) &=& f_-^{(x)}(z|s)
     - {1\over z} B_{--}^{(x+1)}
      \left( A_{+-}^{(x)} f_+^{(x)} + A_{++}^{(x)} f_-^{(x)} \right)(z|s)\ .
   \label{shift:f1}
\end{eqnarray}
Using the identities between operators $A$ and $B$ derived in
appendix~\ref{app:idAB} this can be rewritten as
\begin{eqnarray}
   {1\over\sqrt{z}} f_+^{(x+1)} &=& {1\over z}
      \left( f_+^{(x)}
  +\left( 1-B_{+-}^{(x+1)}\right) A_{++}^{(x)} f_-^{(x)} \right)\ ,
   \nonumber \\
   {1\over\sqrt{z}} f_-^{(x+1)} &=& f_-^{(x)}
     - {1\over z} B_{--}^{(x+1)}
      \left( A_{+-}^{(x)} f_+^{(x)} + A_{++}^{(x)} f_-^{(x)} \right)\ .
   \label{shift:fright}
\end{eqnarray}
Choosing $z=\xi$ or $z=\xib$ here reproduces equations (\ref{result:f}).

Alternatively, one can consider a shift $x+1 \mapsto x$ of the functions
$f_\pm$ which results in the following identities
\begin{eqnarray}
   {1\over\sqrt{z}} f_+^{(x)} &=&
    \left( 1 + {1\over z} B_{++}^{(x)} A_{--}^{(x+1)} \right) f_+^{(x+1)}
     + {1\over z} B_{++}^{(x)} A_{-+}^{(x+1)} f_-^{(x+1)} \ ,
   \nonumber \\
   {1\over\sqrt{z}} f_-^{(x)} &=& {1\over z}
      \left( f_-^{(x+1)}
  +\left( 1+B_{-+}^{(x)}\right) A_{--}^{(x+1)} f_+^{(x+1)} \right)\ .
   \label{shift:fleft}
\end{eqnarray}
Eqs.~(\ref{shift:fright}) and (\ref{shift:fleft}) are equivalent as is
straightforward to show using the identities from appendix~\ref{app:idAB}.
\subsection{Derivatives with respect to $\psi$}
Next, for (\ref{eq:vvdpsi}) and (\ref{B:dpsi}) we take the derivative of
$f_\pm$ with respect to the variable $\psi$ (related to the magnetic field
through Eqs.~(\ref{dresse})). With
\begin{equation}
   {1\over i} {\partial \over \partial \psi} \int_C dz \left( \ldots \right)
      = \xi \left( \ldots \right)_{z=\xi}
      + \xib \left( \ldots \right)_{z=\xib}
\end{equation}
we find from the integral equation (\ref{def:f})
\begin{equation}
  {1\over i} {\partial \over \partial \psi} f_\pm(z|s) =
      -\xi R(z|\xi) f_\pm(\xi|s)
      -\xib R(z|\xib) f_\pm(\xib|s) \; .
   \label{f:dpsi1}
\end{equation}
Using the expression (\ref{rkernelz}) for $R$ in terms of $f_\pm^{(x)}$
this can be expressed as
\begin{eqnarray}
   {\partial \over \partial \psi} f_\pm(z|s) &=&
     {1\over 2\pi}\, \left\{ {\xi\over z-\xi} f_\pm(\xi|s)
              \int_t \left(f_+(\xi|t) f_-(z|t)
                       - f_-(\xi|t)f_+(z|t)\right)
                    \right. \nonumber \\
  && \qquad \left.
      + {\xib\over z-\xib} f_\pm(\xib|s)
              \int_t \left(f_+(\xib|t) f_-(z|t)
                       - f_-(\xib|t)f_+(z|t)\right) \right\}\ .
  \label{f:dpsi2}
\end{eqnarray}
At this point everything follows from the requirement of compatibility
between the shifts (\ref{shift:fright}) or (\ref{shift:fleft}) and the
derivative (\ref{f:dpsi2}). Shifting $\partial_\psi f^{(x)}$ and equating
the result with $\partial_\psi f^{(x+1)}$ one obtains the results announced
in Section~\ref{sec:results}.

On the other hand, taking the $\psi$-derivative of (\ref{def:B}) directly
we obtain
\begin{eqnarray}
  {1\over i} {\partial \over \partial \psi} B_{ab}(s,t) &=&
   {i\over2\pi} \Biggl\{
     f_a(\xi|s) \left[ f_b - \left( B_{+b}\right)^\top f_-
                           + \left( B_{-b}\right)^\top f_+ \right](\xi|t)
   \nonumber \\
   && \qquad
  + f_a(\xib|s) \left[ f_b - \left( B_{+b}\right)^\top f_-
                           + \left( B_{-b}\right)^\top f_+ \right](\xib|t)
      \Biggr\}
\end{eqnarray}
which are Eqs.~(\ref{B:dpsi}) when we identify $F_\pm(s) = f_\pm(\xi|s)$
and $G_\pm(s) = f_\pm(\xib|s)$.

Now we have to compute the logarithmic derivative of our determinant wrt.\
$\psi$:
\begin{equation}
   -i {\partial \over \partial \psi} \ln
            \det\left( 1+\widehat{V} \right) =
      \xi R(\xi|\xi) + \bar{\xi} R(\xib|\xib) \; .
\end{equation}
Starting from (\ref{rkernelz}) we find
\begin{eqnarray}
   \xi R(\xi|\xi) &=& {1\over 2\pi}
        \int_s \left\{ f_+(\xi|s) \partial_\psi f_-(\xi|s)
                    - f_-(\xi|s) \partial_\psi f_+(\xi|s)
              \right\}  \nonumber \\
   && + {1\over 4\pi^2} \; {\xib \over \xi-\xib} \;
        \left(\int_s f_+(\xi|s) f_-(\xib|s)
                    - f_-(\xi|s) f_+(\xib|s) \right)^2 \; .
   \label{eq:rxixi}
\end{eqnarray}
Together with an analogeous expression for $\xib R(\xib|\xib)$ this
completes the proof of the announced result (\ref{eq:vvdpsi}).

\section{The operator-valued Riemann-Hilbert problem
         for the XXX chain
         \label{sec:RH-problem}}
Consider now the integral operator-valued function
\begin{equation}
   M(z) = I + m(z)
   \label{conjgmatrix}
\end{equation}
where
\begin{equation}
   m(z|s,t) = \left( \begin{array}{cc}
               -e_+(z|s) e_-(z|t) & e_+(z|s) e_+(z|t) \\
               -e_-(z|s) e_-(z|t) & e_-(z|s) e_+(z|t) \end{array}
              \right)
\end{equation}
We propose the operator $M(z)$ as the ``conjugation matrix'' for an
infinite dimensional Riemann-Hilbert problem for an integral
operator-valued function $\chi(z)$:
\begin{enumerate}
\item
$\chi(z)$ is analytic outside the contour $C$ (Fig.~\ref{fig:contour}).

\item
$\chi^-(z) = \chi^+(z) \cdot M(z)$ for $z \in C$, and $\chi^\pm$ are the
boundary values of the function $\chi(z)$ as indicated in
Fig.~\ref{fig:contour}.

\item
$\chi(z) \to I$ as $z\to\infty$.
\end{enumerate}
In terms of the corresponding kernels the properties 1--3 can be rewritten
in the following way:
\begin{itemize}
\item[P1.]
$\chi(z|s,t)$ is an analytic function of $z \not\in C$ for all $s,t$.

\item[P2.]
$\chi^-(z|s,t) = \chi^+(z|s,t) + \int_{s'} \chi^+(z|s,s') m(z|s',t)$

\item[P3.]
$ \chi(z|s,t) = \left( \begin{array}{cc} 1 & 0 \\ 0 & 1 \end{array}
                   \right) \delta(s-t) e^t
                 + {1\over z} \Psi_1(s,t) + \ldots$ as $z\to\infty$
\end{itemize}
(As before we suppress the dependence on $x$)

The usual (see Section XV.6 of \cite{vladb}, and \cite{itsx:90})
relationship between the Riemann-Hilbert problem (1--3) and the basic
solutions $f_\pm(z|s)$ of the initial integral operator takes place:
\begin{equation}
   \left( \begin{array}{c} f_+(z|s)\\f_-(z|s) \end{array} \right)
   = \int_t \chi^+(z|s,t)
   \left( \begin{array}{c} e_+(z|t)\\e_-(z|t) \end{array} \right)\ .
   \label{RH:rel}
\end{equation}

Suppose that the solution of the Riemann-Hilbert problem 1--3 exists and is
unique. Then we shall show that the function
\begin{equation}
   \Psi(z) = \chi(z) \cdot \left(
     \begin{array}{cc} z^{-x} & 0 \\ 0 & 1 \end{array} \right)
\end{equation}
satisfies the integral operator-valued linear system (\ref{shift:fright}),
(\ref{f:dpsi2}).

We begin with the $x$-equation. Applying the standard arguments based on
Liouville's theorem and on the $x$ independence of the conjugation integral
operator
\begin{equation}
   M_0(z) = \left( \begin{array}{cc} z^{x} & 0 \\ 0 & 1 \end{array} \right)
       M(z) \left( \begin{array}{cc} z^{-x} & 0 \\ 0 & 1 \end{array} \right)
\end{equation}
we have the equation
\begin{equation}
   \Psi^{(x+1)} \left[ \Psi^{(x)} \right]^{-1} =
   \left( \begin{array}{cc} 0 & 0 \\ 0 & 1 \end{array} \right) \cdot I\
   +\ {1\over z}\ U_0
   \label{Psi:shift}
\end{equation}
where $I(s,t)=\delta(s-t)e^t$ and
\begin{equation}
   U_0 = \chi^{(x+1)} (0) \cdot
   \left( \begin{array}{cc} 1 & 0 \\ 0 & 0 \end{array} \right)
     \cdot \left[ \chi^{(x)}(0) \right]^{-1}
\end{equation}

The operator $m(z)$ is nilpotent $m^2(z)=0$, hence we have
\begin{equation}
   M^{-1}(z) = I - m(z)\ ,
\end{equation}
which in turn implies the equation
\begin{equation}
   \chi^{-1}(z|s,t) = \left(
   \begin{array}{cc} \chi_{22}(z|t,s) & -\chi_{12}(z|t,s) \\
                     -\chi_{21}(z|t,s) & \chi_{11}(z|t,s) \end{array}
   \right)
   \label{chi_inv}
\end{equation}
for the solution of the RH-Problem 1--3. Introducing notations
\begin{equation}
   \chi(0) = \left(
   \begin{array}{cc} 1-B_{+-} &   B_{++} \\
                      -B_{--} & 1+B_{-+} \end{array}
   \right)
   \label{chi0}
\end{equation}
we conclude from (\ref{chi_inv}) that
\begin{equation}
   \chi^{-1}(0) = \left(
   \begin{array}{cc} 1+B_{-+}^\top &  -B_{++}^\top \\
                       B_{--}^\top & 1-B_{+-}^\top \end{array}
   \right)
   \label{chi0_inv}
\end{equation}
(this is an alternative proof of relations (\ref{AB:transpose})!)

Substituting the expansion P3 for $\chi(z)$ as $z\to\infty$ into
(\ref{Psi:shift}), we see that $(U_0)_{11} = 1$. This together with
formulae (\ref{chi0}) and (\ref{chi0_inv}) allow us to rewrite $U_0$ as
\begin{equation}
  U_0 = \left( \begin{array}{cc}
  1 & - \left(1-B_{+-}^{(x+1)}\right)\left(B_{++}^{(x)}\right)^\top \\[8pt]
  - B_{--}^{(x+1)}\left(1+\left(B_{-+}^{(x)}\right)^\top\right)
             & B_{--}^{(x+1)} \left(B_{++}^{(x)}\right)^\top
               \end{array} \right)\ .
\end{equation}

To conclude the analysis of the $x$-equation (\ref{Psi:shift}), one has to
show that the kernels $B_{ab}(s,t)$ from (\ref{chi0}) are just the kernels
introduced in the previous section. To this end, one has to consider the
canonical integral representation for the solution of the problem 1--3
\begin{equation}
  \chi(z_1) = I - {1\over 2\pi i} \int_C {dz_2 \over z_2 - z_1}
           \chi^+(z_2) m(z_2)\ ,
\end{equation}
which implies
\begin{equation}
  \chi(0) = I + {i\over 2\pi} \int_C {dz \over z} \chi^+(z) m(z)\ .
\end{equation}
Taking into account the explicit formula for $m(z)$ and Eq.~(\ref{RH:rel})
we obtain the representations (\ref{def:B}) for $B_{ab}(s,t)$ defined in
(\ref{chi0}).

Eq.~(\ref{Psi:shift}) can be rewritten as
\begin{equation}
   z^{-{1\over2}} \chi^{(x+1)}(z) \cdot \left(
     \begin{array}{cc} z^{-{1\over2}} & 0 \\ 0 & z^{1\over2}\end{array}
     \right) =
   \left[ \left( \begin{array}{cc} 0 & 0 \\ 0 & 1 \end{array} \right)
          \cdot I\ +\ {1\over z}\ U_0 \right] \chi^{(x)}(z)\ .
   \label{chi:shift}
\end{equation}
Applying both sides of (\ref{chi:shift}) to the vector $\left(
\begin{array}{c} e_+^{(x)}(z|s) \\ e_-^{(x)}(z|s) \end{array} \right)$ and
taking into account (\ref{RH:rel}) and that $e_\pm^{(x)} z^{\mp{1\over2}} =
e_\pm^{(x+1)}$ we obtain the basic $x$-equations (\ref{shift:fright}).

\underline{Remark:} The substitution of the expansion P3 into
(\ref{Psi:shift}) yields the following identities for the potentials
$B_{ab}$ (compare with (\ref{AB:inv2})):
\begin{eqnarray}
  \left(1 - B_{+-}^{(x+1)} \right)
      \left( 1 + \left(B_{-+}^{(x)}\right)^\top \right) = 1, &&\qquad
  - \left(1 - B_{+-}^{(x+1)} \right)\left(B_{++}^{(x)}\right)^\top
    = - C_{++}^{(x+1)} \nonumber \\
  - B_{--}^{(x+1)}
    \left( 1 + \left(B_{-+}^{(x)}\right)^\top \right) = -C_{--}^{(x)},
    && \qquad
   B_{--}^{(x+1)}\left(B_{++}^{(x)}\right)^\top
    = C_{-+}^{(x)} - C_{-+}^{(x+1)}\ ,
\end{eqnarray}
where the new potentials $C_{ab}$ are defined as
\begin{equation}
  C_{ab}(s,t) = {i\over 2\pi} \int_C dz f_a(z|s) e_b(z|t), \qquad a,b=\pm\ .
\end{equation}

Let us now study the $\xi$-derivative of the function $\chi(z)$. The
corresponding analysis almost literally repeats the one performed in
\cite{itsx:90} for the Bose-gas case.

We notice (infinite dimensional analogue of the corresponding relation in
\cite{jimbo:80}) that in the neighbourhood of $C$ the function $\chi(z)$
can be represented as
\begin{equation}
   \chi(z) = \widehat{\chi}(z) \cdot \chi_0(z)
   \label{def:chihat}
\end{equation}
where $\widehat{\chi}(z)$ is singlevalued, invertable and analytic in that
neighbourhood, while
\begin{equation}
   \chi_0(z|s,t) = \left[ \delta(s-t)\ e^t\
   -\ {i\over2\pi} \ln{z-\xib\over z-\xi} \left(
   \begin{array}{cc} 0 & 1 \\ 0 & 0 \end{array} \right) \right]\ \cdot\
   \left( \begin{array}{cc} e_-(z|t) & 0 \\
                          - e_-(z|t) & e_+(z|t) \end{array} \right)\ .
\end{equation}
{}From (\ref{def:chihat}) we conclude at once that
\begin{equation}
   \chi_\psi(z) \cdot \chi^{-1}(z) = {A_+ \over z-\xib}
                                   + {A_- \over z-\xi}
   \label{RH:chipsi}
\end{equation}
where
\begin{equation}
  A_\pm = \lim_{z\to z_\pm} \left( z - z_\pm \right)
    \widehat{\chi}(z) \cdot \chi_{0\psi} \cdot \chi_0^{-1}(z)
                      \cdot \widehat{\chi}^{-1}(z)\ , \qquad
    z_+ = \xib\ ,\quad z_- = \xi\ .
\end{equation}
It is easily verified that
\begin{equation}
   \chi_0^{-1}(z|s,t) = \left( \begin{array}{cc}
      e_+(z|s) &0 \\ e_-(z|s) & e_-(z|s) \end{array} \right) \cdot\
   \left[ \delta(s-t)\ e^t\
   +\ {i\over2\pi} \ln{z-\xib\over z-\xi} \left(
   \begin{array}{cc} 0 & 1 \\ 0 & 0 \end{array} \right) \right]\ ,
\end{equation}
which immediately implies
\begin{equation}
   \left( \chi_{0\psi} \chi_0^{-1} \right) (z|s,t) = {1\over2\pi}\left[ {{\xib}
   \label{RH:chi0psi}          \over z-\xib } + {{\xi} \over z-\xi}\right]
                 \left( \begin{array}{cc} 0 & 1 \\ 0 & 0 \end{array} \right)
\end{equation}
Returning to the integral operators $A_\pm$ we obtain from
(\ref{def:chihat})--(\ref{RH:chi0psi}) and (\ref{chi_inv}) that (compare
with the corresponding derivation in \cite{itsx:90}):
\begin{eqnarray}
   A_\pm(s,t) &=& {z_\pm \over 2\pi} \int_{s'} \int_{s''}
      \widehat{\chi}(z_\pm|s,s') \left(
        \begin{array}{cc} 0 & 1 \\ 0 & 0\end{array} \right)
      \widehat{\chi}^{-1}(z_\pm|s'',t)
   \nonumber \\
   &=& {z_\pm \over 2\pi} \int_{s'} \chi(z_\pm|s,s') \left(
     \begin{array}{cc} 0 & e_+(z_\pm|s') \\ 0 & e_-(z_\pm|s') \end{array}
       \right) \cdot \int_{s'} \left(
     \begin{array}{cc} 0 & 0 \\ -e_-(z_\pm|s') & e_+(z_\pm|s') \end{array}
       \right) \chi^{-1}(z_\pm|s',t)
    \nonumber \\
   &=& {z_\pm \over 2\pi} \left(
   \begin{array}{cc}
       - f_+(z_\pm|s) f_-(z_\pm|t) & f_+(z_\pm|s) f_+(z_\pm|t) \\
       - f_-(z_\pm|s) f_-(z_\pm|t) & f_-(z_\pm|s) f_+(z_\pm|t)
   \end{array} \right)\ .
\end{eqnarray}
Remembering the definition of the potentials $F_\pm(s)$, $G_\pm(s)$, we can
rewrite the final formulae for $A_\pm$ as
\begin{eqnarray}
   A_+(s,t) &=& {\xib\over2\pi} \left(
   \begin{array}{cc}
       -G_+(s) G_-(t) & G_+(s)G_+(t) \\
       -G_-(s) G_-(t) & G_-(s)G_+(t)
   \end{array} \right)\ ,
   \nonumber \\
   && \label{RH:Afinal} \\
   A_-(s,t) &=& {\xi\over2\pi} \left(
   \begin{array}{cc}
       -F_+(s) F_-(t) & F_+(s)F_+(t) \\
       -F_-(s) F_-(t) & F_-(s)F_+(t)
   \end{array} \right)\ . \nonumber
\end{eqnarray}
Equations (\ref{RH:chipsi}) and (\ref{RH:Afinal}) imply the basic equation
(\ref{f:dpsi2}) for the $\psi$-derivative of $f_\pm(z|s)$.

\underline{Remark:}
Writing down the compatibility condition for the L--A pair (\ref{Psi:shift}),
(\ref{RH:chipsi}) explicitely
\begin{eqnarray}
   &&\left( {A_+^{(x+1)} \over z-\xib} + {A_-^{(x+1)} \over z-\xi} \right)
   \left( \left(\begin{array}{cc} 0&0\\0&1\end{array}\right)
   +\ {1\over z}\ U_0\right)\
   \nonumber \\
   && \qquad \qquad =\ {1\over z}\ U_{0\psi}\ +\
   \left( \left(\begin{array}{cc} 0&0\\0&1\end{array}\right)
   +\ {1\over z}\ U_0\right)
   \left( {A_+^{(x)} \over z-\xib} + {A_-^{(x)} \over z-\xi} \right)
   \label{RH:compatibility}
\end{eqnarray}
one can obtain some useful identities for the potentials $B_{ab}$, $F_\pm$
and $G_\pm$ that are of course consequences of the basic system
(\ref{result:f}) and (\ref{B:dpsi}).

In a forthcoming publication we intend to use the Riemann-Hilbert problem
1--3 to analyze the asymptotic behaviour of the correlation function
$P(x)$. The results will allow for a comparison of the approach to the
computation of correlation functions used here to the symmetry based one
used e.g.\ in \cite{jimbo:92}.

\section*{Acknowledgements}
H.F. gratefully acknowledges the hospitality at the Institute for
Theoretical Physics in Stony Brook, where much of this work was
performed. This work was partially supported by the Deutsche
For\-schungs\-gemein\-schaft under Grant No.\ Fr~737/2--1 and by the
National Science Foundation (NSF) under Grant No.\ DMS-9315964.

\newpage
\appendix

\section{Completion of the proof of the relation between the Fredholm
         determinants of $\widehat{V}$ and $B_{-+}$
         \label{app:logder}}
To prove the equivalence of Eqs.~(\ref{vvd1}) and (\ref{vvd_ef}) we shall
use the identity
\[
   \ln \det\left(1+{\cal O}\right)
     = \sum_{n>0} \hbox{\rm tr} \left( {\cal O}^n \right)
\]
and consider the equations order by order in $B_{-+}$. The first term is
\begin{equation}
   \hbox{\rm tr} \left( {i\over2\pi} \int_s
           {e_+^{(x)}(z_1|s) f_-^{(x)}(z_2|s) \over z_1} \right)
   = {i\over2\pi} \int_C dz \int_s
           {e_+^{(x)}(z|s) f_-^{(x)}(z|s) \over z}
\end{equation}
which by definition (\ref{def:B}) is equal to
$\hbox{\rm tr}\left(B_{-+}\right)$.
Similarly, to second order we find
\begin{eqnarray}
  && \hbox{\rm tr} \left\{ \left( {i\over2\pi} \int_s
{e_+^{(x)}(z_1|s) f_-^{(x)}(z_2|s) \over z_1} \right)^2 \right\}
  \nonumber \\
  && \qquad = \left({i\over2\pi}\right)^2 \int_C dz_1 \int_C dz_2
       \int_s { {e_+^{(x)}(z_1|s) f_-^{(x)}(z_2|s) \over z_1}}
       \int_t { {e_+^{(x)}(z_2|t) f_-^{(x)}(z_1|t) \over z_2}}
\end{eqnarray}
which, after reordering of terms is equal to
$\hbox{\rm tr}\left\{\left(B_{-+}\right)^2\right\}$. Turning to arbitrary
$n$ now we have ($z_{n+1}\equiv z_1$)
\begin{eqnarray}
  && \left({i\over2\pi}\right)^n
     \left(\prod_\ell \int_c dz_\ell\right)
     \prod_k \int_s
      {e_+^{(x)}(z_k|s_k) f_-^{(x)}(z_{k+1}|s_k) \over z_k}
  \nonumber \\
  && \qquad = \left(\prod_k \int_s\right) \left\{
     \prod_\ell {i\over2\pi} \int_c dz_\ell
       {e_+^{(x)}(z_\ell|s_\ell) f_-^{(x)}(z_{\ell}|s_{\ell-1}) \over z_\ell}
     \right\}
\end{eqnarray}
which completes the proof of (\ref{vvd1}).
\section{Identities between the operators $A_{ab}^{(x)}$ and $B_{ab}^{(x)}$
         \label{app:idAB}}
In this appendix a collection of useful identities in rewriting the
integral equations in this paper is derived. First, from the definition
(\ref{def:A}) of $A_{ab}$ it follows
\begin{eqnarray}
   && A_{+-} \left( 1-B_{+-} \right) - A_{++} B_{--} = 1\ , \qquad
      A_{+-} B_{++} + A_{++} \left( 1+ B_{-+} \right) = 0\ , \nonumber \\
   && A_{--} \left( 1-B_{+-} \right) - A_{-+} B_{--} = 0\ , \qquad
      A_{--} B_{++} + A_{-+} \left( 1+ B_{-+} \right) = 1\ .
   \label{AB:product}
\end{eqnarray}

Next, using the explicit $x$-dependence (\ref{def:e}) of $e_\pm^{(x)}$ and
the shift equation (\ref{shift:f0}) for $f_+$ we find
\begin{eqnarray}
  B_{+-}^{(x+1)}(s,t) &=& {i\over2\pi}
         \int_C {dz\over z} f_+^{(x+1)}(z|s) e_-^{(x)}(z|t) \nonumber \\
  &=& \left( 1- B_{+-}^{(x+1)} \right)
          \left( A_{+-}^{(x)} B_{+-}^{(x)} + A_{++}^{(x)} B_{--}^{(x)}
          \right) \nonumber
\end{eqnarray}
which with (\ref{AB:product}) leads to
\begin{equation}
  \left( 1 - B_{+-}^{(x+1)} \right) A_{+-}^{(x)} = 1\ .
  \label{AB:inv1}
\end{equation}

{}From the definition (\ref{def:B}) we obtain with (\ref{igl:g})
\begin{eqnarray}
   B_{ab}(s,t) &=& {i\over 2\pi}
         \int_C {dz\over z} f_a(z|s) e_b(z|t) \nonumber \\
   &=& {i\over 2\pi}
         \int_C {dz\over z} e_a(z|s) g_b(z|t)\ . \nonumber
\end{eqnarray}
Using the expression (\ref{sol:g}) for $g_\pm$ together with the relations
(\ref{AB:product}) we obtain relations between $B$ and the transpose of
$A$:
\begin{eqnarray}
   && B_{++} = - \left( A_{++} \right)^\top\ , \qquad
      B_{+-} = 1 - \left( A_{-+} \right)^\top\ , \nonumber \\
   && B_{-+} = \left( A_{+-} \right)^\top -1\ , \qquad
      B_{--} = \left( A_{--} \right)^\top\ .
   \label{AB:transpose}
\end{eqnarray}

Finally, we use these identities to get
\begin{equation}
   \left( 1 + B_{-+}^{(x+1)} \right) A_{-+}^{(x)} = 1
  \label{AB:inv2}
\end{equation}
from (\ref{AB:inv1}).


\newpage


\begin{thebibliography}{10}

\bibitem{bethe:31}
H.~Bethe,
\newblock {\em Z. Phys.} {\bf 71}, 205 (1931).

\bibitem{grif:64}
R.~B. Griffiths,
\newblock {\em Phys. Rev.} {\bf 133}, A768 (1964).

\bibitem{vladb}
V.~E. Korepin, A.~G. Izergin, and N.~M. Bogoliubov,
\newblock {\em {Quantum Inverse Scattering Method, Correlation Functions and
  Algebraic Bethe Ansatz}},
\newblock Cambridge University Press, 1993.

\bibitem{fata:book}
L.~D. Faddeev and L.~A. Takhtajan,
\newblock {\em Hamiltonian Methods in the Theory of Solitons},
\newblock Springer Verlag, Berlin, 1987.

\bibitem{ablowitz+segur}
M.~J. Ablowitz and H.~Segur,
\newblock {\em Solitons and the Inverse Scattering Method},
\newblock SIAM, Philadelphia, 1981.

\bibitem{deiftx:93}
P.~A. Deift, A.~R. Its, and X.~Zhou,
\newblock ``Long-time asymptotics for integrable nonlinear wave equations'',
\newblock in {\em Important Developments in Soliton Theory}, edited by A.~S.
  Fokas and V.~E. Zakharov, Springer Verlag, Berlin, 1993.

\bibitem{jimbo:80}
M.~Jimbo, T.~Miwa, Y.~Mori, and M.~Sato,
\newblock {\em Physica} {\bf 1D}, 80 (1980).

\bibitem{korx:94}
V.~E. Korepin, A.~G. Izergin, F.~H.~L. E{\ss}ler, and D.~B. Uglov,
\newblock ``Correlation function of the spin-1/2 {XXX} antiferromagnet'',
  preprint ITP-SB-94-05.

\bibitem{barouch:73}
E.~Barouch, B.~M. McCoy, and T.~T. Wu,
\newblock {\em Phys. Rev. Lett.} {\bf 31}, 1409 (1973).

\bibitem{trmc:73}
C.~A. Tracy and B.~M. McCoy,
\newblock {\em Phys. Rev. Lett.} {\bf 31}, 1500 (1973).

\bibitem{itsx:90}
A.~R. Its, A.~G. Izergin, V.~E. Korepin, and N.~A. Slavnov,
\newblock {\em Int. J. Mod. Phys. B} {\bf 4}, 1003 (1990).

\bibitem{jimbo:92}
M.~Jimbo, K.~Miki, T.~Miwa, and A.~Nakayashiki,
\newblock {\em Phys. Lett. A} {\bf 168}, 256 (1992).

\end{thebibliography}

\newpage

\begin{figure}[h]
\vspace*{2cm}
\begin{center}
\leavevmode
\epsfbox{contour2.eps}
\end{center}
\vspace*{2cm}
\caption{\label{fig:contour}}
Contour $C$ on the unit circle for the integration with respect to $z$ in
the integral operator $\widehat{V}$. '$+$' and '$-$' indicate direction in
which the limit $z\to C$ has to be taken to obtain the boundary values
$\chi^\pm(z)$ for the Riemann-Hilbert problem in Section
\ref{sec:RH-problem}.
\end{figure}

\end{document}